\documentclass[a4paper,12pt]{article}
\usepackage{amsmath}
\usepackage{amssymb}
\usepackage{latexsym}
\usepackage{lscape}
\usepackage{pdflscape}
\newcommand{\be}{\begin{eqnarray}}
\newcommand{\ee}{\end{eqnarray}}
\newcommand{\bs}{\boldsymbol}
\newcommand{\ph}{\phantom}
\newcommand{\bsub}{\begin{subequations}}
\newcommand{\esub}{\end{subequations}}
\newcommand{\non}{\nonumber}
\newcommand{\ket}[1]{|#1\rangle}
\newcommand{\bra}[1]{\langle #1|}

\newcommand{\disfrac}[1][2]{\displaystyle\frac}
\usepackage{hyperref}
\begin{document}
\title{\large\textbf{The General Solution of Bianchi Type $VII_h$ Vacuum Cosmology}}
\author{\textbf{Petros A. Terzis}\thanks{pterzis@phys.uoa.gr} ~\textbf{and} \textbf{T. Christodoulakis}\thanks{tchris@phys.uoa.gr}\\
University of Athens, Physics Department\\
Nuclear \& Particle Physics Section\\
Panepistimioupolis, Ilisia GR 157--71, Athens, Hellas}
\date{}
\maketitle
\begin{center}
\textit{}
\end{center}
\vspace{0.5cm} \numberwithin{equation}{section}
%%%%%%%%%%%%%%%%%%%%%%%%%%%%%%%%%%%%%%%%%%%%%%%%%%%%%%%%%%%%%%%%%%%%%%%%%%%%%%%%%%%%%%%%
\begin{abstract}
The theory of symmetries of systems of coupled, ordinary
differential equations (ODE) is used to develop a concise algorithm
in order to obtain the entire space of solutions to vacuum Bianchi
Einstein’s Field Equations (EFEs). The symmetries used are the well
known automorphisms of the Lie algebra for the corresponding
isometry group of each Bianchi Type, as well as the scaling and the
time re-parametrization symmetry. The application of the method to
Type $VII_h$ results in (a) obtaining the general solution of Type
$VII_0$ with the aid of the third Painlev\'{e} transcendental
$\bs{P}_{III}$; (b) obtaining the general solution of Type $VII_h$
with the aid of the sixth Painlev\'{e} transcendental $\bs{P}_{VI}$
;(c) the recovery of all known solutions (six in total) without a
prior assumption of any extra symmetry; (d) The discovery of a
$\bs{new}$ solution ( the line element given in closed form) with a
$G_3$ isometry group acting on $T_3$, i.e. on time-like
hyper-surfaces, along with the emergence of the line element
describing the flat vacuum Type $VII_0$ Bianchi Cosmology.
\end{abstract}
%%%%%%%%%%%%%%%%%%%%%%%%%%%%%%%%%%%%%%%%%%%%%%%%%%%%%%%%%%%%%%%%%%%%%%%%%%%%%%%%%%%%%%%%
\newpage
\section{Introduction}
The idea of using the group of automorphisms in order to have a
unified development of Bianchi Cosmologies has a long history
\cite{Schuk}. In that direction Harvey \cite{Harvey} was the first
who found the automorphisms of all three-dimensional Lie Algebras,
while the corresponding results for the four-dimensional Lie
Algebras have been reported in \cite{ChrDim}. Jantzen’s tangent
space approach sees the automorphism matrices as the means for
achieving a convenient parametrization of a full scale factor matrix
in terms of a desired, diagonal matrix \cite{jantzen}. Samuel and
Ashtekar were the first to look upon automorphisms from a space
viewpoint \cite{ashtekar}. The notion of \emph{Time-Dependent
Automorphism Inducing Diffeomorphisms} (A.I.D.'s), i.e., coordinate
transformations mixing space and time in the new spatial coordinates
and inducing automorphic motions on the scale-factor matrix, the
lapse, and the shift has been developed in \cite{JMP}. The use of
these covariances enables one to set the shift vector to zero
without destroying manifest spatial homogeneity. At this stage one
can use the ''rigid'' automorphisms, i.e. the remaining ''gauge''
symmetry, as Lie-Point Symmetries of the EFE's in order to reduce
the order of these equations and ultimately completely integrate
them \cite{ChrTer JMP}. The present work of ours consists in the
application of this method to the case of vacuum Bianchi Type
$VII_h$ Cosmology. The method is recapitulated in section 2 while
its application to the above mentioned type, resulting in the
exhaustive discovery of the entire solution space, is given in
section 3. In section 4 we discuss our results and give a brief
description of the solution space in the form of two tables.

\section{The Method}
As it is well known, for spatially homogeneous space-times with a
simply transitive action of the corresponding isometry group
\cite{EM}, \cite{MacBook}, the line element, assumes the form
\begin{equation}\label{line element}
ds^2=\left(N^\alpha N_\alpha-N^2\right)\,dt^2+2\,N_\alpha
\sigma^\alpha_{\ph{a}i}\,dx^idt+
\gamma_{\alpha\beta}\sigma^\alpha_{\ph{a}i}\sigma^\beta_{\ph{a}j}\,dx^idx^j
\end{equation}
where the 1-forms $\sigma^{\alpha}_{i}$, are defined from:
\begin{equation}\label{ορισμός σ}
d\sigma^{\alpha}=C^\alpha_{\ph{a}\beta\gamma}\sigma^\beta\wedge\sigma^\gamma
\Leftrightarrow \sigma^{\alpha}_{\ph{a}i,j} -
\sigma^\alpha_{\ph{a}j,i}=2\,C^\alpha_{\ph{a}\beta\gamma}
\sigma^{\gamma}_{\ph{a}i} \sigma^\beta_{\ph{a}j}.
\end{equation}
Then the field equations are (see e.g. \cite{JMP}):
\begin{equation}\label{τετραγωνικός σύνδεσμος}
E_o\doteq K^{\alpha\beta}K_{\alpha\beta}-K^2- \mathbf{R}=0
\end{equation}
\begin{equation}\label{γραμμικός σύνδεσμος}
E_\alpha\doteq K^\mu_{\ph{a}\alpha}
C^\epsilon_{\ph{a}\mu\epsilon}-K^\mu_{\ph{a}\epsilon}
C^\epsilon_{\ph{a}\alpha\mu}=0
\end{equation}
\begin{equation}\label{δυναμικές εξισώσεις}
E_{\alpha\beta}\doteq
\dot{K}_{\alpha\beta}+N\left(2K^\tau_{\ph{a}\alpha} K_{\tau\beta}-K
K_{\alpha\beta}\right)+2N^\rho\left(K_{\alpha\nu}
C^\nu_{\ph{a}\beta\rho}+K_{\beta\nu} C^\nu_{\ph{a}\alpha\rho}
\right)-N \mathbf{R}_{\alpha\beta}=0
\end{equation}
where
\begin{equation}\label{K}
K_{\alpha\beta}=-\frac{1}{2N}\left(\dot{\gamma}_{\alpha\beta}+2\gamma_{\alpha\nu}
C^\nu_{\ph{a}\beta\rho} N^\rho+2\gamma_{\beta\nu}
C^\nu_{\ph{a}\alpha\rho} N^\rho \right)
\end{equation}
is the extrinsic curvature and
\begin{equation}\label{Rab}
\begin{array}{cc}
\mathbf{R}_{\alpha\beta}=&C^\kappa_{\ph{a}\sigma\tau}
C^\lambda_{\ph{a}\mu\nu}
\gamma_{\alpha\kappa}\gamma_{\beta\lambda}\gamma^{\sigma\nu}\gamma^{\tau\mu}+
2\,C^\kappa_{\ph{a}\beta\lambda} C^\lambda_{\ph{a}\alpha\kappa}+ 2\,
C^\mu_{\ph{a}\alpha\kappa}
C^\nu_{\ph{a}\beta\lambda}\gamma_{\mu\nu}\gamma^{\kappa\lambda}+\\
&\\
 &2\,C^\lambda_{\ph{a}\beta\kappa}
C^\mu_{\ph{a}\mu\nu}\gamma_{\alpha\lambda}\gamma^{\kappa\nu}+ 2\,
C^\lambda_{\ph{a}\alpha\kappa}
C^\mu_{\ph{a}\mu\nu}\gamma_{\beta\lambda}\gamma^{\kappa\nu}
\end{array}
\end{equation}
the Ricci tensor of the hyper-surface.

In \cite{JMP} particular space-time coordinate transformations have
been found, which reveal  as symmetries of (\ref{τετραγωνικός
σύνδεσμος}), (\ref{γραμμικός σύνδεσμος}), (\ref{δυναμικές
εξισώσεις})  the following transformations of the dependent
variables $N, N_\alpha, \gamma_{\alpha\beta}$ :
\begin{equation}\label{gaugetrans}
\tilde{N}=N, \,
\tilde{N}_{\alpha}=\Lambda^\rho_{\ph{a}\alpha}\,(N_\rho+\gamma_{\rho\sigma}\,P^\sigma),
\, \tilde{\gamma}_{\mu\nu}=\Lambda^\alpha_{\ph{a}\mu} \,
\Lambda^\beta_{\ph{a}\nu} \, \gamma_{\alpha\beta}
\end{equation}
where the matrix $\Lambda$ and the triplet $P^\alpha$ must satisfy:
\be\label{intcon}
 \Lambda^\alpha_{\ph{a}\rho} \, C^\rho_{\ph{a}\beta\gamma}& =
&C^\alpha_{\ph{a}\mu\nu}\ \Lambda^\mu_{\ph{a}\beta}
\,\Lambda^\nu_{\ph{a}\gamma}\ee \be\label{intcon1}\
2\,P^\mu\,C^\alpha_{\ph{a}\mu\nu}\Lambda^\nu_{\ph{a}\beta} &=&
\dot{\Lambda}^\alpha_{\ph{a}\beta} \ee

For all Bianchi Types, this system of equations admits solutions
which contain three arbitrary functions of time plus several
constants depending on the Automorphism group of each type. The
three functions of time, are distributed among $\Lambda$ and $P$
(which also contains derivatives of these functions). So one can use
this freedom either to simplify the form of the scale factor matrix
or to set the shift vector to zero. The second action can always be
taken, since, for every Bianchi type, all three functions appear in
$P^\alpha$.

In this work we adopt the latter point of view. When the shift has
been set to zero, there is still a remaining "gauge" freedom
consisting of all constant $\Lambda^\alpha_\beta$ (Automorphism
group matrices). Indeed the system (\ref{intcon}), (\ref{intcon1})
accepts the solution $\Lambda^\alpha_\beta=constant$,
$P^\alpha=\mathbf{0}$. The generators of the corresponding motions
$\tilde{\gamma}_{\mu\nu}=\Lambda^\alpha_{\ph{a}\mu} \,
\Lambda^\beta_{\ph{a}\nu} \, \gamma_{\alpha\beta}$
, induced in the
space of dependent variables spanned by $\gamma_{\alpha\beta}\,'s$
(the lapse is given in terms of
$\gamma_{\alpha\beta},\,\dot{\gamma}_{\alpha\beta}$ by algebraically
solving the quadratic constraint equation ), are  \cite{CMP} :
\begin{equation}\label{genX}
X_{(I)}=\lambda^\rho_{(I)\alpha}\,\gamma_{\rho\beta}\,
\frac{\partial}{\partial\gamma_{\alpha\beta}}
\end{equation}
with $\lambda$ satisfying:
\begin{equation}\label{deflamda}
\lambda^\alpha_{(I)\rho}\,C^\rho_{\ph{a}\beta\gamma}=\lambda^\rho_{(I)\beta}\,
C^\alpha_{\ph{a}\rho\gamma}+
\lambda^\rho_{(I)\gamma}\,C^\alpha_{\ph{a}\beta\rho}.
\end{equation}

Now, these generators define a Lie algebra and each one of them
induces, through its integral curves, a transformation on the
configuration space spanned by the $\gamma_{\alpha\beta}$'s. If a
generator is brought to its normal form (e.g.
$\frac{\partial}{\partial z_{i}}$), then the Einstein equations,
written in terms of the new dependent variables, will not explicitly
involve  $z_{i}$. They thus become a \emph{first order} system in
the function $\dot{z}_{i}$ \cite{Stephani}. If the above Lie algebra
happens to be abelian, then all generators can be brought, to their
normal form simultaneously. If this is not the case, we can
diagonalize in one step the generators corresponding to any eventual
abelian subgroup. The rest of the generators (not brought in their
normal form) continue to define a symmetry of the reduced system of
EFE's if the algebra of the $X_{(I)}$'s is solvable \cite{Olver}.
One can thus repeat the previous step, by choosing one of these
remaining generators. This choice will of course depend upon the
simplifications brought to the system at the previous level. Finally
if the algebra does not contain any abelian subgroup, one can always
choose one of the generators, bring it to its normal form, reduce
the system and search for its symmetries (if there are any). Lastly,
two further symmetries of (\ref{τετραγωνικός σύνδεσμος}),
(\ref{γραμμικός σύνδεσμος}), (\ref{δυναμικές εξισώσεις}) are also
present and can be used in conjunction with the constant
automorphisms: The time reparameterization $t \rightarrow
f(t)+\alpha$, owing to the non-explicit appearance of time in these
equations, and the scaling by a constant $\gamma_{\alpha\beta}
\rightarrow \mu \gamma_{\alpha\beta}$ as can be straightforwardly
verified. Their corresponding generators are: \\\be\label{ορισμός
Y1} Y_1  =\frac{1}{\dot{f}}\, \frac{\partial}{\partial t} \ee
\begin{equation}\label{ορισμός Y2}
Y_2=\gamma_{\alpha\beta}\,
\frac{\partial}{\partial\gamma_{\alpha\beta}}
\end{equation}

These generators commute among themselves, as well as with the
$X_{(I)}$'s, as it can be easily checked.

\section{Application to Bianchi Type $VII_h$}
We are now going to apply the Method, previously discussed, to the
case of Bianchi Type $VII_h$. For this type the structures constants
are
\begin{equation}\label{σταθερές δομής}
\begin{array}{lll}
C^1_{\ph{1}13}=-C^1_{\ph{1}31}=C^2_{\ph{2}23}=-C^2_{\ph{2}32}=-h\\
C^1_{\ph{1}32}=-C^1_{\ph{1}23}=C^2_{\ph{2}13}=-C^2_{\ph{2}31}=1\\
C^\alpha_{\ph{a}\beta\gamma}=0 & for\, all\, other\, values\, of\,
\alpha \beta \gamma
\end{array}
\end{equation}
Using these values in the defining relation (\ref{ορισμός σ}) of the
1-forms $\sigma^{\alpha}_{i}$ we obtain
\begin{equation}
\sigma^\alpha_i=\left(\begin{matrix} 0 & e^{h\,x}\,\sin x &
e^{h\,x}\,\cos x \cr 0 & e^{h\,x}\,\cos x & -e^{h\,x}\,\sin x \cr
\frac{1}{2} & 0 & 0
\end{matrix}
\right)
\end{equation}
The corresponding vector fields $\xi^i_\alpha$ (satisfying
$\left[\xi_\alpha,\xi_\beta\right]=\frac{1}{2}\,C^\gamma_{\ph{a}\alpha\beta}
\xi_\gamma$) with respect to which the Lie Derivative of the above
1-forms is zero are: \be\label{killing}
\begin{array}{lll}
\xi_1=\partial_y & \xi_2=\partial_z &
\xi_3=\partial_x+(z-h\,y)\partial_y-(y+h\,z)\partial_z
\end{array}
\ee

The Time Depended A.I.D.'s are described by
\begin{equation}\label{Aut Λ}
 \Lambda^\alpha_{\ph{a}\beta}=
\left(\begin{matrix} c\,e^{h\,P(t)}\cos P(t) & c\, e^{h\,P(t)}\sin
P(t) & x(t) \cr
  -c\,e^{hP(t)}\sin P(t) & c\,e^{h\,P(t)}\cos P(t) & y(t)
  \cr
  0 & 0 & 1\end{matrix}
 \right)
\end{equation} and
\be \label{Aut P} P^\alpha
&=&\left(\frac{x(t)\dot{P}(t)+h^{2}\,x(t)\dot{P}(t)-h\,\dot{x}(t)+\dot{y}(t)}{2\,(1+h^{2})},
\right. \\ &&\left.
\frac{y(t)\dot{P}(t)+h^{2}\,y(t)\dot{P}(t)-h\,\dot{y}(t)-\dot{x}(t)}{2\,(1+h^{2})},
\frac{\dot{P}(t)}{2}\right) \ee where $P(t), x(t)$ and $y(t)$ are
arbitrary functions of time. As we have already remarked the three
arbitrary functions appear in $P^\alpha$ and thus can be used to set
the shift vector to zero.

The remaining symmetry of the EFE's is, consequently, described by
the constant matrix:
\begin{equation}\label{Outer Aut}
M=\left(\begin{matrix}e^{s_{1}}& s_2 &s_{3} \cr -s_2
&e^{s_{1}}&s_{4} \cr 0&0&1\end{matrix}\right)
\end{equation}
where the parametrization has been chosen so that the matrix becomes
identity for the zero value of all parameters.

 Thus the induced transformation on the scale factor matrix is
$\tilde{\gamma}_{\alpha\beta}=M^{\mu}_{\alpha}M^{\nu}_{\beta}\gamma_{\mu\nu}$,
which explicitly reads:
\begin{equation}\label{gamma new}
\left\{
\begin{array}{l}
\tilde{\gamma}_{11}=e^{2\,{s_1}}\,{{\gamma }_{11}}-2\,e^{s_1}\,s_2\,\gamma_{12}+s_2^2\,\gamma_{22}\\\
\\
\tilde{\gamma}_{12}=e^{{2\,s_1}}\,{{\gamma }_{12}}-s_2^2\,\gamma_{12}+e^{s_1}\,s_2\,(\gamma_{11}-\gamma_{22})\\
\\
\tilde{\gamma}_{13}=e^{{s_1}}\,\left( {s_3}\,{{\gamma }_{11}} + {s_4}\,{{\gamma }_{12}} + {{\gamma }_{13}}\right)
-s_2\,\left(s_3\,\gamma_{12}+s_4\,\gamma_{22}+\gamma_{23}\right)\\
\\
\tilde{\gamma}_{22}=e^{2\,s_1}\,\gamma_{22}+2\,e^{s_1}\,s_2\,\gamma_{12}+s_2^2\,\gamma_{11}\\
\\
\tilde{\gamma}_{23}=e^{{s_1}}\,\left( {s_3}\,{{\gamma }_{12}} +
{s_4}\,{{\gamma }_{22}} + {{\gamma }_{23}} \right)+
s_2\,\left(s_3\,\gamma_{11}+s_4\,\gamma_{12}+\gamma_{13}\right)\\
\\
\tilde{\gamma}_{33}={{s_3}}^2\,{{\gamma }_{11}} + 2\,{s_3}\,
   \left( {s_4}\,{{\gamma }_{12}} + {{\gamma }_{13}} \right)  + {{s_4}}^2\,{{\gamma }_{22}} +
  2\,{s_4}\,{{\gamma }_{23}} + {{\gamma }_{33}}
\end{array}\right.
\end{equation}

The previous equations, define a group of transformations $G_{r}$ of
dimension  $r=dim(Aut(VII_h))=4$. The four generators of the group,
can be evaluated from the relation:
\begin{equation}\label{ορισμός Χ}
X_{A}=\left(\frac{\partial\tilde{\gamma}_{\alpha\beta}}{\partial
s_{A}}\right)_{s=0}\frac{\partial}{\partial\gamma_{\alpha\beta}}
\end{equation}
where $A=\left\{1,2,3,4\right\}$. Applying this definition to
(\ref{gamma new}) we have the generators:
\begin{equation}\label{X1}
X_{1}=2\,\gamma_{11}\frac{\partial}{\partial\gamma_{11}}+2\,\gamma_{12}\frac{\partial}{\partial\gamma_{12}}
+\gamma_{13}\frac{\partial}{\partial\gamma_{13}}+2\,\gamma_{22}\frac{\partial}{\gamma_{22}}+
\gamma_{23}\frac{\partial}{\gamma_{23}}
\end{equation}
\begin{equation}\label{X2}
X_{2}=-2\,\gamma_{12}\frac{\partial}{\partial\gamma_{11}}
+\left(\gamma_{11}-\gamma_{22}\right)\frac{\partial}{\partial\gamma_{12}}
-\gamma_{23}\frac{\partial}{\partial\gamma_{13}}
+2\,\gamma_{12}\frac{\partial}{\partial\gamma_{22}}
+\gamma_{13}\frac{\partial}{\partial\gamma_{23}}
\end{equation}
\begin{equation}\label{X3}
X_{3}=\gamma_{11}\frac{\partial}{\partial\gamma_{13}}
+\gamma_{12}\frac{\partial}{\partial\gamma_{23}}
+2\,\gamma_{13}\frac{\partial}{\partial\gamma_{33}}
\end{equation}
\begin{equation}\label{X4}
X_{4}=\gamma_{12}\frac{\partial}{\partial\gamma_{13}}
+\gamma_{22}\frac{\partial}{\partial\gamma_{23}}
+2\,\gamma_{23}\frac{\partial}{\partial\gamma_{33}}
\end{equation}

The algebra $\textsl{g}_{r}$ that corresponds to the group $G_{r}$
has the following table of commutators:
\begin{equation}\label{μεταθέτες}
\begin{array}{lll}
\left[X_{1},X_{2}\right]=0,&\left[X_{1},X_{3}\right]=X_3,&\left[X_{1},X_{4}\right]=X_{4},
\\ \left[X_{2},X_{3}\right]=-X_{4},& \left[X_{2},X_{4}\right]=X_3,&\left[X_{3},X_{4}\right]=0
\end{array}
\end{equation}

As it is evident from the above commutators (\ref{μεταθέτες})  the
group is non-abelian, so we cannot diagonalize  at the same time all
the generators. However, if we calculate the derived algebra of
$\textsl{g}_{r}$, we have
\begin{equation}
\textsl{g}_{r'}=\left\{[X_{A},X_{B}]: X_{A}, X_{B}\in
\textsl{g}_{r}\right\}\Rightarrow
\textsl{g}_{r'}=\left\{X_{3},X_{4}\right\}
\end{equation}
and furthermore, it's second derived algebra reads:
\begin{equation}
\textsl{g}_{r''}=\left\{[X_{A},X_{B}]: X_{A}, X_{B}\in
\textsl{g}_{r'}\right\}\Rightarrow \textsl{g}_{r''}=\left\{0\right\}
\end{equation}

Thus, the group  $G_{r}$ is solvable since the $\textsl{g}_{r''}$ is
zero. As it is evident  $X_{3}, X_{4}, Y_{2}$ generate an Abelian
subgroup, and we can, therefore,  bring them to their normal form
simultaneously. The appropriate transformation of the dependent
variables is: \be\label{gammav} \left\{
\begin{array}{l}
\gamma_{11}= e^{{u_1} -{u_6}} \\
\\
\gamma_{12}=e^{{u_1} -{u_6}}\,u_2\\
\\
\gamma_{13}=e^{{u_1} - {u_6}}\,\left(u_3 + u_2\,u_5 \right) \\
\\
\gamma_{22}=e^{{u_1}-{u_6}}\,u_4\\
\\
\gamma_{23}= e^{{u_1} - {u_6}}\,\left( u_2\,u_3 +u_4\,u_5 \right) \\
\\
\gamma_{33}=e^{u_1-u_6}\,\left(e^{u_6}+{{u_3}}^2 + 2\,u_2\,u_3\,u_5
+ u_4\,u_5^2 \right)
\end{array}\right.
\ee

In these coordinates the generators $Y_{2}, X_{A}$  assume the form:

\be\label{generatorsv}
\begin{array}{lll}
Y_2=\frac{\partial}{\partial u_1} \quad X_4=\frac{\partial}{\partial
u_5} \quad X_3=\frac{\partial}{\partial u_3}  \\
\\
X_2=\left(1+2\,u_2^2-u_4\right)\,\frac{\partial}{\partial u_2}
-u_5\,\frac{\partial}{\partial
u_3}+2\,\left(u_2+u_2\,u_4\right)\,\frac{\partial}{\partial
u_4}+u_3\,\frac{\partial}{\partial
u_5}+2\,u_2\,\frac{\partial}{\partial u_6} \\
\\
X_1 =-u_3\,\frac{\partial}{\partial u_3}-u_5
\frac{\partial}{\partial u_5}-2\,\frac{\partial}{\partial u_6}
\end{array}
\ee

Evidently, a first look at (\ref{gammav}) gives the feeling that it
would be hopeless even to write down the Einstein equations.
However, the simple form of the first three of the generators
(\ref{generatorsv}) ensures us that these equations will be of first
order in the functions $\dot{u}_1$,  $\dot{u}_3$ and $\dot{u}_5$.
\subsection{Description of the Solution Space}

Before we begin solving the Einstein equations, a few comments on
the allowable range of values for the functions $u_{i},
i={1,\ldots,6}$ will prove very useful.

The determinant of  $\gamma_{\alpha\beta}$, is
\begin{equation}\label{detgamma}
det[\gamma_{\alpha\beta}]=e^{3\,u_1 - 2\,u_6}\,\left(-u_2^2+
u_4\right)
\end{equation}
so we must have $u_4\,>\,u_2^2$ .

The two linear constraint equations, written in the new variables
(\ref{gammav}), give \be E_1=0 \Rightarrow
\frac{1}{2}\,e^{-u_6}\,\left( (3\,h-u_2)\,\dot{u_3}+(3\,h\,u_2-u_4)\,\dot{u_5} \right)  =0 \\
E_2=0\Rightarrow \frac{1}{2}\,e^{-u_6}\,\left(
(1+3\,h\,u_2)\,\dot{u_3}+(u_2+3\,h\,u_4)\,\dot{u_5}\right)=0 \ee
This system admits only the trivial solution, since the determinant
of the $2\times2$ matrix formed by the coefficients of $\dot{u}_3,
\dot{u}_5$ becomes zero only for the forbidden value $u_4=u_2^2$. We
thus have \be u_3=k_3, & u_5=k_5 \ee Now, these values of $u_3,u_5$
make $\gamma_{13},\gamma_{23}$ functionally dependent upon
$\gamma_{11},\gamma_{12},\gamma_{22}$ (see (\ref{gammav})). It is
thus possible to set these two components to zero by means of an
appropriate constant automorphism.

\emph{We therefore can, without loss of generality, start our
investigation of the solution space for Type $VII_h$ vacuum Bianchi
Cosmology from a block-diagonal form of the scale-factor matrix
(and, of course, zero shift)} \be\label{gammau1}
\gamma_{\alpha\beta}=\left(
\begin{matrix}
\gamma_{11} & \gamma_{12} & 0 \cr \gamma_{12} & \gamma_{22} & 0 \cr
0 & 0 & \gamma_{33}
\end{matrix}
\right) \ee These unknown functions of time have to satisfy the
quadratic and the third linear constraint, as well as the spatial
EFE's. As we have earlier remarked, since the algebra
(\ref{μεταθέτες}) is solvable, the remaining (reduced) generators
$X_1,X_2$ (corresponding to block-diagonal constant automorphisms)
as well as $Y_2$ continue to define a Lie-Point symmetry of the
reduced EFE's and can thus be used for further integration of this
system of equations.

The remaining (reduced) automorphism generators are \be
X_{1}=2\,\gamma_{11}\,\frac{\partial}{\partial\gamma_{11}}
+2\,\gamma_{12}\,\frac{\partial}{\partial\gamma_{12}}
+2\,\gamma_{22}\,\frac{\partial}{\partial\gamma_{22}}\\
X_{2}=-2\,\gamma_{12}\,\frac{\partial}{\partial\gamma_{11}}
+\left(\gamma_{11}-\gamma_{22}\right)\,\frac{\partial}{\partial\gamma_{12}}
+2\,\gamma_{12}\,\frac{\partial}{\partial\gamma_{22}} \nonumber \ee
The appropriate change of dependent variables which brings these
generators -along with $Y_2$- into normal form, is described by the
following scale-factor matrix :\be\label{gammaured}
\gamma_{\alpha\beta}=\left(
\begin{matrix}
\frac{1}{2}\,e^{u_1+2\,u_6}\,\left(1-2\,u_2\,\sin 2\,u_4\right) &
e^{u_1+2\,u_6}\,u_2\,\cos 2\,u_4& 0 \cr e^{u_1+2\,u_6}\,u_2\,\cos
2\,u_4 & \frac{1}{2}\,e^{u_1+2\,u_6}\,\left(1+2\,u_2\,\sin
2\,u_4\right) & 0 \cr 0 & 0 & e^{u_1}
\end{matrix} \right)
\ee The generators are now reduced to \be
Y_2=-\frac{\partial}{\partial u_1}, \, X_2=\frac{\partial}{\partial
u_4}, \, X_1=-\frac{\partial}{\partial u_6} \ee indicating that the
system will be of first order in the derivatives of these variables.
The remaining variable $u_2$ will enter, (along with
$\dot{u}_2,\,\ddot{u}_2$) explicitly in the system and is therefore
advisable (if not mandatory) to be used as the time parameter, i.e.
to effect the change of time coordinate \be t\rightarrow u_2(t)=s,
\, u_1(t)\rightarrow u_1(t(s)), \, u_4(t)\rightarrow u_4(t(s)), \,
u_6(t)\rightarrow u_6(t(s)). \ee This choice of time will of course
be valid only if $u_2$ is not a constant. We are thus led to
consider two cases according to the constancy or non-constancy of
this dependent variable.

Until now, we haven't commented upon the range of values that the
parameter $h$ can attain. As it is well known, for the value $h=0$
we come across the \emph{Class A} model, which admits a Lagrangian
description, whereas for $h\neq 0$ we have the \emph{Class B} model
which lacks such a Lagrangian description. So we are forced to
examine two further possibilities, as to whether $h$ is equal to, or
different from, zero.

\subsubsection{Case I: $h\,=\,0$ and $u_2(t)\,=\,k_2$}
In the parametrization (\ref{gammaured}) the determinant of
$\gamma_{\alpha\beta}$, is
\begin{equation*}
det[\gamma_{\alpha\beta}]=\frac{1}{4}\,e^{3\,u_1 + 4\,u_6}\,\left(1-
4\,k_2^2\right)
\end{equation*}
so we must have $-\frac{1}{2}<k_2<\frac{1}{2}$ . The third linear
constraint reads \be E_3=0\Rightarrow
\frac{8\,k_2^2\,\dot{u_4}}{-1+4\,k_2^2}=0\Rightarrow u_4=k_4 & or
&k_2=0 \ee The case $u_4=k_4$ leads, through equation $E_{34}=0$ to
$k_2=0$ . Thus, the only possibility is $k_2=0$. Substituting this
value into the quadratic constraint equation $E_0$ we obtain \be
-\frac{1}{2}\,\left(3\,\dot{u_1}^2+8\,\dot{u_1}\,\dot{u_6}+4\,\dot{u_6}^2\right)=0
\ee which has the following two solutions \bsub\label{eq Eo}
\be\label{a} u_1=k_1-2\,u_6 \\ \label{b} u_1=k_1-\frac{2}{3}\,u_6\ee
\esub For the first of \eqref{eq Eo} all the spatial EFE's are
equivalent to the equation  \be
2\,\dot{u}_6\,N\dot{N}+2\,N^2\,\left(\dot{u}_6^2-\ddot{u}_6\right)=0
\ee from which we have for the lapse function \be
N^2=k\,e^{-2\,u_6}\,\dot{u}_6^2\ee

Choosing a time  parametrization
$u_6=-\frac{1}{2}\,\ln(\frac{\tau^2}{k})$, and using the
automorphism matrix  \eqref{Outer Aut} with entries
$s_1=\frac{1}{2}\,\left(\ln2-k_1\right),\, s_2=s_3=s_4=0$ we arrive
at the line element \be\label{flat
h=0}d\,s^2=-d\,\tau^2+\tau^2\,d\,x^2+d\,y^2+d\,z^2 \ee which
describes a \emph{flat} space admitting a manifest $VII_0$ symmetry
\cite{Taub}. To the best of our knowledge, it is the first time that
this line element emerges in the course of investigation of the
solution space to this Bianchi Type.

For the second of \eqref{eq Eo} all the spatial EFE's are equivalent
to the equation \be
2\,\dot{u}_6\,N\dot{N}-2\,N^2\,\left(\dot{u}_6^2+\ddot{u}_6\right)
=0\ee which gives the lapse function \be
N^2=k\,e^{2\,u_6}\,\dot{u}_6^2\ee

Choosing a time  parametrization
$u_6=\frac{1}{2}\,\ln(\frac{4\,\tau^3}{9\,k})$, redefining the
constant $k_1=\frac{1}{3}\,\ln\frac{256}{9\,k}$  and using the
automorphism matrix \eqref{Outer Aut} with entries
$s_1=\frac{1}{2}\,\ln\frac{9\,k}{8},\, s_2=s_3=s_4=0$ we arrive at
the line element \be\label{diagonal
h=0}d\,s^2=-\tau\,d\,\tau^2+\frac{1}{\tau}\,d\,x^2+\tau^2\,d\,y^2
+\tau^2\,d\,z^2 \ee

This line element was first derived by Ellis
\cite{Ellis},\cite{Stewart Ellis} and admits, besides the three
killing fields \eqref{killing} (with $h=0$),  a fourth symmetry
generator\be\label{4th killing} \xi_4=\partial_x \ee along with a
homothetic vector field \be\label{hom diagonal}
\eta=2\,\tau\,\partial_\tau+4\,x\,\partial_x+y\,\partial_y+z\,\partial_z
\ee There is thus a $G_4$ symmetry group acting (of course, multiply
transitively) on each $V_3$ of this metric. However, it is
interesting to note that we have not imposed the extra symmetry from
the beginning, but rather it emerged as a result of the
investigation process.

\subsubsection{Case II: $h\,=\,0$ and $u_2(t)\,=\,t$}
With this choice of time gauge the third linear constraint reads \be
E_3=0\Rightarrow\frac{8\,t^2}{4\,t^2-1}\,\dot{u}_4=0 \Rightarrow
u_4=k_4 \ee With this information at hand, the quadratic constraint
$E_o$ yields the lapse function \be\label{lapse case II} N^2 =
-\frac{e^{u_1}}{64\,t^2}\,\left((4\,t^2-1)\,(\dot{u}_1+2\,
\dot{u_6})\,(3\,\dot{u}_1+2\,
\dot{u_6})+16\,t\,(\dot{u}_1+\dot{u_6})+4\right)\ee We now turn to
the spatial equations of motion and substitute the above lapse. The
simplest is $E_{33}=0$ and the coefficient of $\ddot{u}_1$ in this
equation is proportional to the quantity
\begin{equation*}\left((4\,t^2-1)\,\dot{u}_6+2\,t\right)\,\dot{u}_1
+(4\,t^2-1)\,\dot{u}_6^2+4\,t\,\dot{u}_6+1 \end{equation*} which can
be safely regarded different from zero, since by setting this
quantity equal to zero and solving for $\dot{u}_1$ we end up with
zero lapse (with the help of the rest of the equations of motion).
We can thus solve $E_{33} = 0$ for $\ddot{u}_1$ and substitute into
$E_{11} = 0$. In this transformed equation $E_{11} = 0$, the
coefficient of $\ddot{u}_6$ is proportional to
\begin{equation*}
(-2\,t+\sin(2\,k_4))\,(\dot{u}_1+\dot{u}_6)-1 \end{equation*} a
quantity which is different from zero, since it's nihilism leads
again to zero lapse. From the transformed $E_{11}=0$ we have the
expression for $\ddot{u}_6$, so we finally arrive a the following
polynomial system of first order in $\dot{u}_1,\, \dot{u}_6$
\be\label{u1'',
u6''}\ddot{u}_1=\bra{\dot{u}_1}\,A_1\,\ket{\dot{u}_6}, &
\ddot{u}_6=\bra{\dot{u}_1}\,A_2\,\ket{\dot{u}_6}\ee where we have
introduced the notation $\bra{\dot{u}_i}=\left( 1 \, \dot{u}_i \,
\dot{u}_i^2\, \dot{u}_i^3 \right)$ and
$\ket{\dot{u}_i}=\bra{\dot{u}_i}^t$ with the $4\times4$ matrices
$A_1,\,A_2$ given by \be\label{A1A2} A_1=\begin{pmatrix}
\frac{4}{4\,t^2-1} &\frac{16\,t}{4\,t^2-1} & 4 & 0\\
\frac{4\,t^2+1}{t\,(4\,t^2-1)} & 4
&\frac{-4\,t^2-1}{t}& 0 \\ -1& \frac{2\,(-4\,t^2+1)}{t}& 0&0 \\
\frac{3\,(1-4\,t^2)}{4\,t} & 0 & 0 &0 \end{pmatrix},\,
A_2=\begin{pmatrix} \frac{6}{-4\,t^2+1} &
\frac{-28\,t^2+1}{t\,(4\,t^2-1)}&-8 &\frac{-4\,t^2+1}{t}\\
\frac{16\,t}{-4\,t^2+1} & -12 & \frac{-8\,t^2+2}{t}&0\\
-3 & \frac{-12\,t^2+3}{4\,t}& 0 & 0 \\
0&0&0&0\end{pmatrix}\nonumber\\\ee Due to the form of $A_1, A_2$
(their components are rational functions of the time $t$), system
\eqref{u1'', u6''} can be partially integrated with the help of the
following Lie-B\"{a}cklund transformation
\begin{equation}\label{u1', u6'}
\begin{split}
\dot{u}_1(t) &
=\frac{-16\,r^2(t)+1}{4\,(4\,t^2-1)\,r(t)}-4\,t\,\dot{r}(t) \\
\\\dot{u}_6(t)&=\frac{-16\,r^2(t)+16\,t\,r(t)+3}{8\,(-4\,t^2+1)\,r(t)}+2\,t\,\dot{r}(t)
\end{split} \end{equation}
resulting in the single, second order ODE for the function $r(t)$
\be\label{equation r} \ddot{r}=
-\frac{1}{r}\,\dot{r}^2+\frac{(12\,t^2+1)\,r+t}{(-4\,t^2+1)\,t\,r}\,\dot{r}
\ee At this stage, in order to solve \eqref{equation r} we apply the
contact transformation: \be\label{contact}\begin{split} r(t)& =
\frac{\xi\,w'(\xi)}{4\,w(\xi)}\,,&t & =
\frac{1}{2}-\frac{1}{1+w(\xi)}\,,& \dot{r}(t)&
=-\frac{(w(\xi)-1)(w(\xi)+1)^3}{8\,w'(\xi)\,w(\xi)}
\end{split}\ee which reduces it to
\be\label{eq w}
w''(\xi)=\frac{w'(\xi)^2}{w(\xi)}-\frac{w'(\xi)}{\xi}
-\frac{1}{2}\,\frac{w(\xi)^2-1}{\xi}
 \ee which is nothing else but the third Painlev\'e transcendent
$w:=\mathbf{P_{III}}(\alpha,\beta,\gamma,\delta)$ with entries
$(\alpha,\beta,\gamma,\delta)=(-1/2,1/2,0,0)$. For completeness we
give the general form of the equation that the third Painlev\'e
transcendent satisfies: \be\label{Pain III}
w''(\xi)=\frac{w'(\xi)^2}{w(\xi)}-\frac{w'(\xi)}{\xi}+
\frac{\alpha\,w(\xi)^2+\beta}{\xi}+\gamma\,w(\xi)^3+
\frac{\delta}{w(\xi)}\ee Using the final equation \eqref{eq w}, the
contact transformation \eqref{contact} and the Lie-B\"{a}cklund
transformation \eqref{u1', u6'} we find that the functions $u_6$ and
$u_1'$ are given by \bsub\label{u6, u1'}\be\label{u6a}
u_6(\xi)&=\disfrac{1}{4}\,
\ln\left(\Big|\disfrac{\xi}{w(\xi)}\Big|(w(\xi)+1)^2\right)-\disfrac{1}{2}\,u_1(\xi)\\
\label{u1'b}u_1'(\xi)&=\disfrac{\xi\,w'(\xi)^2}{4\,w(\xi)^2}
+\disfrac{1}{4}\,w(\xi)+
\disfrac{1}{4\,w(\xi)}-\frac{1}{4\,\xi}-\frac{1}{2} \ee\esub and the
lapse function has the form \be\label{final lapse h=0}
N^2=\frac{1}{16\,\xi}\,e^{u_1} \ee The scale-factor matrix
$\gamma_{\alpha\beta}$ is thus \be\label{final γ h=0}
\gamma_{\alpha\beta}=\begin{pmatrix}
\disfrac{1}{2}\,\sqrt{\Big|\disfrac{\xi}{w(\xi)}\Big|}\,(w(\xi)+1) &
\disfrac{1}{2}\,\sqrt{\Big|\disfrac{\xi}{w(\xi)}\Big|}\,(w(\xi)-1)
&0
\\ \disfrac{1}{2}\,\sqrt{\Big|\disfrac{\xi}{w(\xi)}\Big|}\,(w(\xi)-1)
&\disfrac{1}{2}\,\sqrt{\Big|\disfrac{\xi}{w(\xi)}\Big|}\,(w(\xi)+1)
& 0 \\ 0&0& e^{u_1(\xi)} \end{pmatrix} \ee which can be brought to
diagonal form with the aid of the automorphism matrix \eqref{Outer
Aut} with entries $s_1=-\frac{1}{2}\,\ln2, s_2=-\frac{1}{\sqrt{2}}$
\be\gamma_{\alpha\beta}=\begin{pmatrix}
\sqrt{\Big|\disfrac{\xi}{w(\xi)}\Big|}\,w(\xi) & 0 &0
\\ 0
&\sqrt{\Big|\disfrac{\xi}{w(\xi)}\Big|} & 0 \\ 0&0& e^{u_1(\xi)}
\end{pmatrix} \ee Gathering all the pieces  we arrive at
the final form of the line element \be\label{final h=0}
d\,s^2&=&\kappa^2\,\Big(-\frac{1}{16\,\xi}\,e^{u_1}\,d\,\xi^2+\frac{1}{4}\,e^{u_1}\,d\,x^2
+\sqrt{\Big|\frac{\xi}{w}\Big|}\,\left(w\,\sin^2x+\cos^2x\right)\,d\,y^2
\nonumber \\
&&+\sqrt{\Big|\frac{\xi}{w}\Big|}\,\sin(2\,x)\,\left(w-1\right)\,d\,y\,d\,z
+\sqrt{\Big|\frac{\xi}{w}\Big|}\,\left(w\,\cos^2x+\sin^2x\right)\,d\,z^2\Big)
\ee which represents the \emph{general} solution of Bianchi Type
$VII_0$ non-flat Vacuum Cosmology, since it contains the expected
number of three essential constants (two implicit in the third
Painlev\'e transcendent plus the overall $\kappa$). The above line
element was first given by Lorenz-Petzold \cite{Lorenz}, but it was
not then pointed out that it represented the general solution.
\\ \\
\textbf{Particular Solutions}

In order for the contact transformation \eqref{contact} to be well
defined, it is obvious that the function $w(\xi)$ must not be
constant. However, remarkably enough, the resulting line element
(\ref{final h=0}) \emph{does not} inherit this restriction. Thus, if
there is some constant solution to equation \eqref{eq w}, it could
produce a particular solution through \eqref{final h=0}. By
inspection it is obvious that \eqref{eq w} admits the solutions
$w(\xi)=\pm 1$, so we could use them to obtain two particular
solutions.
\\ \\
$\blacktriangleright$ \textbf{Subcase $w(\xi)=1$}

With this value of $w(\xi)$ \eqref{final h=0} indicates that
$\xi>0$, so plugging this value to \eqref{u6, u1'} we have \be
u_1(\xi)=k_1-\frac{1}{4}\,\ln\xi\, &
u_6(\xi)=-\frac{1}{2}\,k_1+\frac{1}{8}\,\ln(16\,\xi^3) \ee which,
after using the usual simplifications brought by the automorphism
matrix \eqref{Outer Aut} and redefining the variable $\xi$ to
$\xi=\tau^4$, results in  \be
d\,s^2=-\tau\,d\,\tau^2+\frac{1}{\tau}\,d\,x^2+\tau^2\,d\,y^2
+\tau^2\,d\,z^2 \ee which is the line element \eqref{diagonal h=0}.
\\ \\
$\blacktriangleright$ \textbf{Subcase $w(\xi)=-1$}

Now from \eqref{final h=0} we must have $\xi<0$ and from
\eqref{u1'b} we obtain \be
u_1(\xi)=k_1-\xi-\frac{1}{4}\,\ln\big|\xi\big| \ee while from
\eqref{u6a} $u_6$ remains undefined. The line element \eqref{final
h=0} with the help of the automorphism matrix \eqref{Outer Aut} and
the definition $\xi=-\tau^4$ becomes \be\label{T3}
d\,s^2=\kappa^2\,\Big(e^{\tau^4}\,\tau\,d\,\tau^2+
\frac{1}{4\,\tau}\,e^{\tau^4}\,d\,x^2+\tau^2\,\cos(2\,x)\,d\,y^2-\nonumber
\\
2\,\tau^2\,\sin(2\,x)\,d\,y\,d\,z-\tau^2\,\cos(2\,x)\,d\,z^2\Big)
\ee which, though physically acceptable, corresponds to Bianchi Type
$VII_0$ symmetry on $T_3$, and  was first given by Barnes
\cite{Barnes}.

\subsubsection{Case III: $h\,\neq\,0$ and $u_2(t)\,=\,k_2$}
In this case the determinant of the scale factor matrix
$\gamma_{\alpha\beta}$ is \be
det(\gamma_{\alpha\beta})=\frac{1}{4}\,e^{3\,u_1+4\,u_6}\,(1-4\,k_2^2)\ee
so we must have $-\frac{1}{2}<k_2<\frac{1}{2}$ in order for it to be
positively defined.

The third linear constraint $E_3=0$ reads \be \frac{1}{4\,k_2^2-1}\,
\left(8\,k_2^2\,\dot{u_4}+2\,h\,(1-4\,k_2^2)\,\dot{u}_6\right)=0\Rightarrow
u_6=k_6+\frac{4\,k_2^2}{h\,(4\,k_2^2-1)}\,u_4\ee and the quadratic
constraint $E_o=0$ gives for the lapse function $N^2$ \be
N^2=\frac{e^{u_1}}{16\,h^2\,(4\,k_2^2-1)\,(3\,h^2\,(4\,k_2^2-1)-4\,k_2^2)}
\,\Big(3\,h^2\,(4\,k_2^2-1)^2\,\dot{u}_1^2+\non\\
\non\\ 32\,h\,k_2^2\,(4\,k_2^2-1)\,\dot{u_1}\,\dot{u}_4
+16\,k_2^2\,(h^2\,(4\,k_2^2-1)+4\,k_2^2)\,\dot{u}_4^2\Big) \ee

Now we are ready to attack the spatial equations of motion after
substituting in them the above lapse. $E_{33}=0$ is again the
simplest one. In this equation, the coefficient of $\ddot{u}_4$ is
proportional to \be\label{pro term}
k_2^2\,\dot{u}_1\,\left(h\,(4\,k_2^2-1)\,\dot{u}_1
+(4\,k_2^2+h^2\,(4\,k_2^2-1))\,\dot{u}_4\right) \ee so in order to
solve $E_{33}=0$ for $\ddot{u}_4$ we must ensure that the above
quantity is different from zero. Setting this quantity equal to zero
we get \bsub\label{three case} \be\label{th a}
u_1&=&k_1-\frac{4\,k_2^2+h^2\,(4\,k_2^2-1)}{h\,(4\,k_2^2-1)}\,u_4
\\\label{th b} u_1&=&k_1 \\ \label{th c} k_2&=&0 \ee \esub

$\blacktriangleright$ The solution \eqref{th a} leads to
inconsistency.

$\blacktriangleright$ The solution \eqref{th b} forces equation
$E_{33}=0$ to give either $k_2=\pm\disfrac{h}{2\,\sqrt{h^2+1}}$
which leads to zero lapse or $k_2=\pm\disfrac{h}{2\,\sqrt{h^2-1}}$,
which makes the determinant of $\gamma_{\alpha\beta}$ negative so is
unacceptable.

$\blacktriangleright$ The solution \eqref{th c} satisfies all the
spatial equations and leads, after the usual simplifications
achieved by the automorphism matrix \eqref{Outer Aut} and the choice
of time gauge $u_1=2\,\ln(2\,h\,\tau)$, to the line element
\be\label{flat h}
d\,s^2=-d\,\tau^2+h^2\,\tau^2\,d\,x^2+e^{2\,h\,x}\,\tau^2\,d\,y^2
+e^{2\,h\,x}\,\tau^2\,d\,z^2 \ee which describes a \emph{flat} space
admitting a manifest $VII_h$ symmetry, a line-element first
presented by Doroshkevich \emph{et al} \cite{Dor} and reproduced by
Siklos \cite{Siklos}.

Having ensured that the term \eqref{pro term} is not equal to zero
we can solve $E_{33}=0$ for $\ddot{u}_4$ and substitute into the
other equations of motion. From $E_{11}=0$ we have
\begin{multline}(\sin(2\,u_4)-2\,k_2)\,
\left(\dot{u}_1+\frac{2\,(3\,h^2\,(4\,k_2^2-1)+4\,k_2^2)}{3\,h\,(4\,k_2^2-1)}\,\dot{u}_4\right)
\\
\left(\dot{u}_1-\frac{2\,(4\,k_2^2\,(h^2-1)-h^2)}{h\,(4\,k_2^2-1)}\,\dot{u}_4\right)
=0 \end{multline} which leads to the following possibilities
\bsub\label{pos} \be \label{pos a}
u_4&=&\frac{1}{2}\,\arcsin(2\,k_2)\\\label{pos b}
u_1&=&k_1-\frac{2\,(3\,h^2\,(4\,k_2^2-1)+4\,k_2^2)}{3\,h\,(4\,k_2^2-1)}\,u_4
\\ \label{pos c}
u_1&=&k_1+\frac{2\,(4\,k_2^2\,(h^2-1)-h^2)}{h\,(4\,k_2^2-1)}\,u_4
\ee\esub

$\blacktriangleright$ The solution \eqref{pos a} leads to zero
lapse.

$\blacktriangleright$ The solution \eqref{pos b} leads to
inconsistency.

$\blacktriangleright$ The solution \eqref{pos c} satisfies all the
spatial equations and leads, after the usual simplifications with
the automorphism matrix \eqref{Outer Aut} and the choice of time
gauge $u_1=\tau$, to the line element \begin{multline}\label{G6
metric}
d\,s^2=\frac{1}{4}\,e^{\frac{-2\,\lambda^2+2\,h^2\,(\lambda^2-1)}{h\,(\lambda^2-1)}\,\tau}
\left(-d\,\tau^2+d\,x^2\right)-2\,\lambda\,e^{2\,h\,(\tau+x)}\,\sin2(\tau+x)\,d\,y\,d\,z+
\\ e^{2\,h\,(\tau+x)}\,\bigg(1+\lambda\,\cos2(\tau+x)\bigg)\,d\,y^2
+e^{2\,h\,(\tau+x)}\,\bigg(1-\lambda\,\cos2(\tau+x)\bigg)\,d\,z^2
\end{multline}
which was presented in \cite{Dor} and \cite{Siklos}.
This line element admits, besides the three killing fields
\eqref{killing}, three more, namely \bsub\label{3 new kill}
\be\label{kil a} \xi_4&=&
e^{\frac{-\lambda^2+h^2\,(\lambda^2-1)}{h\,(\lambda^2-1)}\,(x-\tau)}\,\partial_\tau
-e^{\frac{-\lambda^2+h^2\,(\lambda^2-1)}{h\,(\lambda^2-1)}\,(x-\tau)}\,\partial_x
\\ \non \\ \label{kil b} \xi_5&=&
y\,e^{\frac{-\lambda^2+h^2\,(\lambda^2-1)}{h\,(\lambda^2-1)}\,(x-\tau)}\,\partial_\tau
-y\,e^{\frac{-\lambda^2+h^2\,(\lambda^2-1)}{h\,(\lambda^2-1)}\,(x-\tau)}\,
\partial_x\non \\
&&+\,e^{\frac{-\lambda^2-h^2\,(\lambda^2-1)}{h\,(\lambda^2-1)}\,(x+\tau)}
\big(c_1\,\cos2\,(\tau+x)+c_2\,\sin2\,(\tau+x)+c_3\big)\,\partial_y \non \\
&&+\,e^{\frac{-\lambda^2-h^2\,(\lambda^2-1)}{h\,(\lambda^2-1)}\,(x+\tau)}
\big(c_2\,\cos2\,(\tau+x)-c_1\,\sin2\,(\tau+x)\big)\,\partial_z
\\ \non \\\label{kil c} \xi_6&=&
z\,e^{\frac{-\lambda^2+h^2\,(\lambda^2-1)}{h\,(\lambda^2-1)}\,(x-\tau)}\,\partial_\tau
-z\,e^{\frac{-\lambda^2+h^2\,(\lambda^2-1)}{h\,(\lambda^2-1)}\,(x-\tau)}\,
\partial_x\non \\
&&+\,e^{\frac{-\lambda^2-h^2\,(\lambda^2-1)}{h\,(\lambda^2-1)}\,(x+\tau)}
\big(c_2\,\cos2\,(\tau+x)-c_1\,\sin2\,(\tau+x)\big)\,\partial_y \non \\
&&-\,e^{\frac{-\lambda^2-h^2\,(\lambda^2-1)}{h\,(\lambda^2-1)}\,(x+\tau)}
\big(c_1\,\cos2\,(\tau+x)+c_2\,\sin2\,(\tau+x)-c_3\big)\,\partial_z\ee\esub
where the constants $(c_1,c_2,c_3)$ are given by
\bsub\label{constants c} \be\label{c1} c_1&=&
-\frac{\lambda\,h\,\big(\lambda^2+h^2\,(\lambda^2-1)\big)}
{4\,\big(\lambda^4+h^4\,(\lambda^2-1)^2+2\,h^2\,(\lambda^2-1)(3\,\lambda^2-2)\big)}
\\ \label{c2} c_2&=&
\frac{\lambda\,h^2\,(\lambda^2-1)}
{2\,\big(\lambda^4+h^4\,(\lambda^2-1)^2+2\,h^2\,(\lambda^2-1)(3\,\lambda^2-2)\big)}
\\ \label{c3}
c_3&=&\frac{h}{4\,\big(\lambda^2+h^2\,(\lambda^2-1)\big)}\ee\esub
Again it is worth mentioning that this $G_6$ symmetry was not
imposed from the begging but emerged during the seeking of the
solution space. The non-vanishing commutators are
\begin{align}\label{comm G6}
\left[ \xi_1,\xi_5 \right]&=\xi_4  &\left[\xi_2,\xi_6\right] &= \xi_4 \non \\
\left[\xi_3,\xi_4\right]&=2\,(\frac{c_1}{c_2}+h)\,\xi_4 &
\left[\xi_3,\xi_5\right]&=2\,(\frac{c_1}{c_2}+\frac{h}{2})\,\xi_5+\xi_6
\non \\
\left[\xi_3,\xi_6\right]&=-\xi_5+2\,(\frac{c_1}{c_2}+\frac{h}{2})\,\xi_6
& &
\end{align}
with $(c_1,c_2)$ given by \eqref{constants c}. Finally the line
element \eqref{G6 metric} admits  a homothetic vector field
\be\label{hom G6}
\eta=\frac{h\,(\lambda^2-1)}{-\lambda^2+h^2\,(\lambda^2-1)}\,\partial_\tau-
\frac{h\,(\lambda^2-1)}{-\lambda^2+h^2\,(\lambda^2-1)}\,\partial_x
+y\,\partial_y+z\,\partial_z\ee

\subsubsection{Case IV: $h\,\neq\,0$ and $u_2(t)\,=\,t$}
In this case the determinant of the sale factor matrix becomes \be
det[\gamma_{\alpha\beta}]=\frac{1}{4}\,e^{3\,u_1+4\,u_6}\,(1-4\,t^2)
\non\ee so we must demand that $|t|\leq\disfrac{1}{2}$ in order for
$\gamma_{\alpha\beta}$ to be positive definite.

The third linear constraint $E_3=0$ can be used to define the
function $u_6$ \be E_3=0\Rightarrow
\dot{u}_6=\frac{2\,t\,(2\,t\,\dot{u}_4-h)}{h\,(4\,t^2-1)}
\Rightarrow u_6=k_6+\int
\frac{2\,t\,(2\,t\,\dot{u}_4-h)}{h\,(4\,t^2-1)}\,d\,t \ee The
quadratic constraint $E_o=0$ defines the lapse function $N^2$
\be\label{lapse
III}N^2=\frac{e^{u_1}}{16\,h^2\,(4\,t^2-1)\,(3\,h^2\,(4\,t^2-1)-4\,t^2)}
\,\Big(3\,h^2\,(4\,t^2-1)^2\,\dot{u}_1^2+\non\\
\non\\ 32\,h\,t^2\,(4\,t^2-1)\,\dot{u_1}\,\dot{u}_4
+16\,t^2\,(h^2\,(4\,t^2-1)+4\,t^2)\,\dot{u}_4^2-4\,h^2\Big) \ee
Substituting the above values of the lapse $N^2$ and the function
$u_6$ in equation $E_{33}=0$ we find the coefficient of $\ddot{u}_1$
to be proportional to\be
4\,t^2\,\left(4\,t^2+h^2\,(4\,t^2-1)\right)\,\dot{u}_4^2+
4\,h\,t^2\,(4\,t^2-1)\,\dot{u}_1\,\dot{u}_4 -h^2\non\ee a quantity
that can be safely regarded different from zero, since it's nihilism
leads either to zero lapse or to inconsistency. Thus we can solve
$E_{33}=0$ for $\ddot{u}_1$ and substitute it to $E_{11}=0$. In
order to solve this equation for $\ddot{u}_4$ we must be assured
that it's coefficient does not vanish. Setting this coefficient
equal to zero we arrive at the following equation \be
\dot{u}_1=\frac{h\,\cos 2\,u_4+2\,t}{2\,t\,(\sin 2\,u_4-2\,t)}-
\frac{(4\,t^2\,(h^2+1)-h^2)\,\dot{u}_4-2\,h\,t}{h\,(4\,t^2-1)}\non\ee
which is unacceptable  because it leads to inconsistency. After
solving equation $E_{11}=0$ for $\ddot{u}_4$ we finally arrive to
the following polynomial system of first order in
$\dot{u}_1,\,\dot{u}_4$ \be\label{ddu1,
ddu4}\ddot{u}_1=\bra{\dot{u}_1}\,B_1\,\ket{\dot{u}_4}, &
\ddot{u}_4=\bra{\dot{u}_1}\,B_2\,\ket{\dot{u}_4}\ee where we have
used  again the notation $\bra{\dot{u}_i}=\left( 1 \, \dot{u}_i \,
\dot{u}_i^2\, \dot{u}_i^3 \right)$ and
$\ket{\dot{u}_i}=\bra{\dot{u}_i}^t$ with the $4\times4$ matrices
$B_1,\,B_2$ given by \be\label{B1B2} B_1&=f\begin{pmatrix}
\frac{4\,( h^2 - 4\,( -1 + h^2 ) \,t^2 ) }{{( 1 - 4\,t^2 ) }^4} & 0
& \frac{16\,t^2\,( -16\,t^4 + h^4\,{( 1 - 4\,t^2 ) }^2 ) }
  {h^2\,{( 1 - 4\,t^2 ) }^4} & 0 \\ \frac{-4\,t\,( 1 - 8\,t^2 + 6\,h^2\,( -1 + 4\,t^2 )  ) }
  {{( -1 + 4\,t^2 ) }^3} & \frac{32\,t^2\,( -h^2 + 4\,( -1 + h^2 ) \,t^2 ) }
  {h\,{( -1 + 4\,t^2 ) }^3} & \frac{-4\,t\,( -h^2 + 4\,( -1 + h^2 ) \,t^2 ) \,
    ( 4\,t^2 + 3\,h^2\,( -1 + 4\,t^2 )  ) }{h^2\,
    {( -1 + 4\,t^2 ) }^3}& 0 \\ \frac{-3\,h^2 + 12\,( -1 + h^2 ) \,t^2}{{( 1 - 4\,t^2 ) }^2}& \frac{32\,t^3}{h\,{( 1 - 4\,t^2 ) }^2}& 0& 0\\ \frac{3\,t}{-1 + 4\,t^2}& 0& 0 & 0\end{pmatrix}\nonumber \\ \non \\
B_2&=g\begin{pmatrix} \frac{2\,h}{4\,t^2 + h^2\,( 1 - 4\,t^2 ) } & \frac{-6\,h^2\,{( 1 - 4\,t^2 ) }^2 + 4\,t^2\,( -3 + 8\,t^2 ) }
  {-4\,t^3 + h^2\,t\,( -1 + 4\,t^2 ) } & \frac{8\,( 4\,t^4 + h^2\,t^2\,( -1 + 4\,t^2 )  ) }{-4\,h\,t^2 + h^3\,( -1 + 4\,t^2 ) } & \frac{-4\,t\,( -3\,h^2 + 4\,t^2 + 12\,h^2\,t^2 ) }{h^2} \\ 0 & \frac{16\,t^2\,( -1 + 4\,t^2 ) }{-h^2 + 4\,( -1 + h^2 ) \,t^2} & \frac{32\,t^3\,( -1 + 4\,t^2 ) }{-4\,h\,t^2 + h^3\,( -1 + 4\,t^2 ) } & 0 \\ \frac{3\,h\,{( 1 - 4\,t^2 ) }^2}{-8\,t^2 + h^2\,( -2 + 8\,t^2 ) } & \frac{3\,t\,{( 1 - 4\,t^2 ) }^2}{-h^2 + 4\,( -1 + h^2 ) \,t^2} & 0 &0 \\ 0 &0&0&0 \nonumber
\end{pmatrix} \non \ee  where \be f=\frac{3\,h^2\,(4\,t^2-1)-4\,t^2}{(4\,t^2-1)^2}, & g =\disfrac{(4\,t^2-1)\,(3\,h^2\,(4\,t^2-1)-4\,t^2)}{4\,(h^2-1)\,t^2-h^2} \ee Due to the form of $B_1, B_2$
(their components are rational functions of the time $t$), system
\eqref{ddu1, ddu4} can be partially integrated with the help of the
following Lie-B\"{a}cklund transformation \bsub\label{du1, du4} \be
\dot{u}_1&=&\frac{h^2\,(3\,h^2\,(4\,t^2-1)+ 4\,t^2)\,\tan
r(t)}{2\,h\,t\,\sqrt{1-4\,t^2}\,(3\,h^2\,(4\,t^2-1)-4\,t^2)}
-\frac{4\,(h^2-1)\,t^2-h^2}{2\,h\,\sqrt{1-4\,t^2}}\,\dot{r}(t) \\
\non\\ \dot{u}_4&=&\frac{3\,h^2\,\sqrt{1-4\,t^2}\,\tan
r(t)}{4\,t\,(3\,h^2\,(4\,t^2-1)-4\,t^2)}
+\frac{1}{4}\,\sqrt{1-4\,t^2}\,\dot{r}(t) \ee\esub yielding  the
single second order ODE for the function $r(t)$ \be\label{equation r
II} \ddot{r}&=&\left(\frac{h}{2}\,\sqrt{1-4\,t^2}+\tan
r\right)\,\dot{r}^2  +\non\\&&
\left(-\frac{3\,(4\,t^2\,(3\,h^2-1)-h^2)}{t\,(4\,(3\,h^2-1)\,t^2-3\,h^2)}
+\frac{h\,(3\,h^2\,(4\,t^2-1)-8\,t^2)}{t\,(4\,(3\,h^2-1)\,t^2-3\,h^2)\,\sqrt{1-4\,t^2}}
\,\tan r\right)\,\dot{r}+ \non\\&& \frac{9\,h^4\,(\sin
2\,r+h\,\sqrt{1-4\,t^2})}{2\,t^2\,(4\,t^2
+3\,h^2\,(1-4\,t^2))^2}\,\sec^2
r-\frac{h\,(3\,h^2(1-4\,t^2)+8\,t^2)^2}{2\,t^2\,(3\,h^2\,(1-4\,t^2)+4\,t^2)^2
\,\sqrt{(1-4\,t^2)^3}} \non \\ \ee This equation contains all the
information concerning the unknown part of the solution space of the
Type $VII_h$ vacuum Cosmology ($h\neq0$). Unfortunately, it does not
posses any Lie-point symmetries that can be used to reduce its order
and ultimately solve it. However, its form can be substantially
simplified through the use of new dependent and independent variable
$(\rho,u(\rho))$ according to  $
r(s)=\pm\arcsin{\frac{u(\rho)}{\sqrt{\rho^2-1}}},\,
s=\sqrt{\frac{3\,h^2\,(\rho-1)}{12\,h^2\,(\rho-1)+8}} $ thereby
obtaining the equation \be\label{final u III}
  \ddot{u}=\pm\frac{h\,(1-\dot{u}^2)}{\sqrt{(6\,h^2\,\rho+4-6\,h^2)\,(\rho^2-u^2-1)}}
  \Rightarrow \ddot{u}^2=\frac{h^2\,(1-\dot{u}^2)^2}{(6\,h^2\,
  \rho+4-6\,h^2)\,(\rho^2-u^2-1)}
  \ee

This equation is a special case of the general equation\be\label{equ
general} \ddot{u}^2=\frac{(1-\dot{u}^2)^2}{(\kappa+
\lambda\,\rho)\,(\rho^2-u^2-1)}\ee with the values
$\kappa=-6+\disfrac{4}{h^2},\, \lambda=6$. The general solution of
\eqref{equ general} was first given in  \cite{ChrTer CQG} and can be
obtained as follows: First we apply the contact transformation:
\be\label{contact II}
\begin{split}
u(\rho)& =
-\frac{8}{\lambda}\,y(\xi)+\frac{4\,(2\,\xi-1)}{\lambda}\,y'(\xi) &\rho & =  -\frac{\kappa}{\lambda}+\frac{4}{\lambda}\,y'(\xi)\\
\dot{u}(\rho)& =  2\,\xi-1 & \ddot{u}(\rho)& = \frac{\lambda}{2\,y''(\xi)}
\end{split}\ee
which reduces it to \be\label{y equation} \xi^2\,(\xi-1)^2\,{y''}^2=
-4y'\,(\xi\,y'-y)^2+4\,{y'}^2\,(\xi\,y'-y)-\frac{\kappa}{2}\,{y'}^2+
\frac{\kappa^2-\lambda^2}{16}\,y'
 \ee

This equation is a special form of the equation SD-Ia, appearing in \cite{Cosgrove}, where  a classification of second order second
degree ordinary differential equations was performed. The general
solution of (\ref{y equation}) is obtained with the help of the
sixth Painlev\'{e} transcendent
$w:=\mathbf{P_{VI}}(\alpha,\beta,\gamma,\delta)$ and reads:
\be\label{solution y} y & = &
\frac{\xi^2\,(\xi-1)^2}{4\,w\,(w-1)(w-\xi)}\,\left(w'-\frac{w\,(w-1)}{\xi\,(\xi-1)}\right)^2
\nonumber \\ & & +\frac{1}{8}\,(1\pm
\sqrt{2\,\alpha})^2\,(1-2\,w)-\frac{\beta}{4}\,\left(1-\frac{2\,\xi}{w}\right)
\nonumber \\ & &
-\frac{\gamma}{4}\,\left(1-\frac{2\,(\xi-1)}{w-1}\right)+ \left(\frac{1}{8}-\frac{\delta}{4}\right)
\, \left(1-\frac{2\,\xi\,(w-1)}{w-\xi}\right) \ee where the sixth
Painlev\'{e} transcendent
$w:=\mathbf{P_{VI}}(\alpha,\beta,\gamma,\delta)$ is defined by the
ODE:

\be\label{Painleve 6} w'' & = & \frac{1}{2}\left( \frac{1}{w-1} +
\frac{1}{w} + \frac{1}{w-\xi } \right) \,{w'}^2 -\left( \frac{1}{\xi-1} + \frac{1}{\xi} + \frac{1}{w-\xi}
\right) \,w' \nonumber \\
& & +\frac{w\,\left( w-1 \right) \,\left( w-\xi\right) }
  {{\xi^2\,\left( \xi-1 \right) }^2} \left(\alpha +\beta\,
  \frac{\xi}{{w}^2}  + \gamma\, \frac{\left(\xi-1 \right)}
  {{\left( w-1 \right) }^2} +\delta\,\frac{\xi\,\left(\xi -1\right) }
   {{\left( w-\xi\right) }^2}\right) \ee

 The values of the parameters
$\left(\alpha,\beta,\gamma,\delta\right)$ of the Painlev\'{e}
transcendent, can be obtained from the solution of the following
system: \bsub\label{system} \be \alpha-\beta+\gamma-\delta \pm
\sqrt{2\,\alpha}+1 & =&-\frac{\kappa}{2} \\
\left(\beta+\gamma\right)\,\left(\alpha+\delta \pm \sqrt{2\,\alpha}\right) &=&0 \\
\left(\gamma-\beta\right)\,\left(\alpha-\delta \pm
\sqrt{2\,\alpha}+1\right)+\frac{1}{4}\,\left(\alpha-\beta-\gamma+\delta
\pm \sqrt{2\,\alpha}\right)^2 & = & \frac{\kappa^2-\lambda^2}{16} \\
\frac{1}{4}\,\left(\gamma-\beta\right)\,\left(\alpha+\delta \pm
\sqrt{2\,\alpha}\right)^2+\frac{1}{4}\,\left(\beta+\gamma\right)^2\,\left(\alpha-\delta
\pm \sqrt{2\,\alpha}+1\right) & = & 0 \ee \esub

Plugging in (\ref{system}) the values of $\kappa=-6+\disfrac{4}{h^2}, \lambda=6$ for
Type $VII_h$, we have twenty-four solutions (counting multiplicities) of this system. In order for the parameters
$(\alpha,\beta,\gamma,\delta)$ to be real numbers we end up only with four possibilities
\bsub\label{par}\be (\alpha,\beta,\gamma,\delta)&=& \left(\disfrac{4\,h^2-1}{2\,h^2}-\sqrt{3-\disfrac{1}{h^2}},\disfrac{1}{2\,h^2},-\disfrac{1}{2\,h^2}, \disfrac{1-2\,h^2}{2\,h^2}\right)  \non\\
(\alpha,\beta,\gamma,\delta)&=& \left(\disfrac{4\,h^2-1}{2\,h^2}+\sqrt{3-\disfrac{1}{h^2}},\disfrac{1}{2\,h^2},-\disfrac{1}{2\,h^2}, \disfrac{1-2\,h^2}{2\,h^2}\right) ,\, |h| \geq \disfrac{1}{\sqrt{3}}\\ \textrm{and} \non\\
(\alpha,\beta,\gamma,\delta)&=& \left(\disfrac{1}{2},\disfrac{2-3\,h^2}{2\,h^2}+\disfrac{\sqrt{1-3\,h^2}}{h^2}, \disfrac{3\,h^2-2}{2\,h^2}+\disfrac{\sqrt{1-3\,h^2}}{h^2},\disfrac{1}{2}\right)\non \\(\alpha,\beta,\gamma,\delta)&=& \left(\disfrac{1}{2},\disfrac{2-3\,h^2}{2\,h^2}-\disfrac{\sqrt{1-3\,h^2}}{h^2}, \disfrac{3\,h^2-2}{2\,h^2}-\disfrac{\sqrt{1-3\,h^2}}{h^2},\disfrac{1}{2}\right),\, |h| \leq \disfrac{1}{\sqrt{3}}\non \\\ee\esub
 For the values $h=\pm\disfrac{1}{\sqrt{3}}$ the above relations coincide and as we will show these values of $h$ give rise to a particular solution.

Gathering all the pieces the final form of the \emph{general} line
element describing  Bianchi Type $VII_h$ vacuum Cosmology is
\be\label{final VII}
d\,s^2&=&\kappa^2\,\Big(-\frac{e^{u_1(\xi)}}{16\,h^2\,\xi\,(\xi-1)}\,(\bs{d}\,\xi)^2+
\sqrt{\xi\,(\xi-1)}\left(\sqrt{y'(\xi)}-
\sin(2\,u_4(\xi))\,\sqrt{y'(\xi)-\frac{1}{h^2}}\right)\,(\bs{\sigma}^1)^2
\non \\
&&+2\,\cos(2\,u_4(\xi))\,\sqrt{\xi\,(\xi-1)\,(y'(\xi)-\frac{1}{h^2})}\,\bs{\sigma}^1\,\bs{\sigma}^2
\non \\&&+ \sqrt{\xi\,(\xi-1)}\left(\sqrt{y'(\xi)}+
\sin(2\,u_4(\xi))\,\sqrt{y'(\xi)-\frac{1}{h^2}}\right)\,(\bs{\sigma}^2)^2+
e^{u_1(\xi)}\,(\bs{\sigma}^3)^2 \Big)\ee where \bsub\label{u1 and
u4}\be u_1'(\xi) & = & \frac{\left( -1 + h^2 \right) \,\left( -1 +
2\,\xi \right)  + 2\,h^2\,y(\xi )}
  {2\,h^2\,\left( -1 + \xi  \right) \,\xi } \ee \\
\be u_4'(\xi) & = & \frac{1 - 2\,\xi  + 2\,h^2\,y(\xi )}
  {4\,h\,\left( -1 + \xi  \right) \,\xi \,\left( -1 + h^2\,y'(\xi ) \right) }\ee \esub
and $ y(\xi)$ is given by \eqref{solution y}. Again, this line
element contains three essential constants, thus representing the
general solution to the EFE's for the Class B $VII_h$ case.
\\ \\
\textbf{Particular Solutions}

Even though the line element \eqref{final VII} represents the
general solution of Bianchi Type $VII_h$ vacuum Cosmology, it does
not come into a manageable form due to the appearance of the sixth
Painlev\'{e} transcendent.  To partially remedy this inconvenience,
we give, in the following, some closed form line-elements arising
from particular solutions to \eqref{solution y} and \eqref{Painleve
6}.
\\ \\
$\blacktriangleright$ \textbf{Subcase $y(\xi)=c$ and $|h|\leq \disfrac{1}{\sqrt{3}}$}

One way to obtain a particular solution from the above line element
\eqref{final VII} is to follow the reasoning of Case II, i.e to
observe that, although the form of the contact transformation
\eqref{contact II} implies that the function $y(\xi)$  cannot be
constant, the line element \eqref{final VII} is free of this
restriction; the difficulty with the negative argument in the square
root is circumvented by using the hyperbolic sine/cosine (see
\eqref{T3 VIIh} below). We can thus check if the assumption
$y(\xi)\equiv c$ leads to a particular solution. Skipping the
calculational details, we find that for $y(\xi)\equiv
c=\frac{\sqrt{1-3\,h^2}}{2\,h^2}$ all the Einstein's field equations
are satisfied and we end up with the $\textbf{new}$ line element
\be\label{T3 VIIh} d\,s^2&=&\kappa^2\,\sin(4\,h\,\tau) \Big(f(\tau)\,(\bs{d}\,\tau)^2 +\sin(h\,\ln f(\tau))\,(\bs{\sigma}^1)^2- \sin(h\,\ln f(\tau))\,(\bs{\sigma}^2)^2 \non \\
&&+ 2\,\cos(h\,\ln f(\tau))\,\bs{\sigma}^1\,\bs{\sigma}^2 +f(\tau)\, (\bs{\sigma}^3)^2\Big)   \\
 \non f(\tau)& = & \sin^{-\frac{1}{h^2}}(4\,h\,\tau)\, \tan^{-\frac{\sqrt{1-3\,h^2}}{h^2}}(2\,h\,\tau),\qquad |h|\leq \frac{1}{\sqrt{3}}\ee which even though is physically acceptable it corresponds to Bianchi Type $VII_h$ symmetry on $T_3$.
 Since the above line element admits only the three killing fields \eqref{killing} and no homothetic vector field we can
 conclude that the constant $\kappa$ is essential.

 An interesting property of the line element \eqref{T3 VIIh} is that, for the value $h^2=\disfrac{1}{3}$, i.e.
 \be\label{G4 h^2=1/3} d\,s^2&=& \kappa^2\Big(\csc^2\frac{4\,\tau}{\sqrt{3}}(\bs{d}\,\tau)^2 -\sin\frac{4\,\tau}{\sqrt{3}}\,\sin\big(\sqrt{3}\,\ln\sin\frac{4\,\tau}{\sqrt{3}}\big)\,(\bs{\sigma}^1)^2 \non\\ &&+\,\sin\frac{4\,\tau}{\sqrt{3}}\,\sin\big(\sqrt{3}\,\ln\sin\frac{4\,\tau}{\sqrt{3}}\big)\,(\bs{\sigma}^2)^2\non \\
 && +2\,\sin\frac{4\,\tau}{\sqrt{3}}\,\cos\big(\sqrt{3}\,\ln\sin\frac{4\,\tau}{\sqrt{3}}\big)\,\bs{\sigma}^1\,\bs{\sigma}^2 +\csc^2\frac{4\,\tau}{\sqrt{3}}(\bs{\sigma}^3)^2\Big)\ee  admits a fourth killing field, namely
 \be \eta=e^\frac{-2\,x}{\sqrt{3}}\,\sin\frac{4\,\tau}{\sqrt{3}}\,\partial_\tau-2\,e^\frac{-2\,x}{\sqrt{3}}\,\cos\frac{4\,\tau}{\sqrt{3}}\,\partial_x \ee The geometry \eqref{G4 h^2=1/3} was first given by Petrov \cite{Petrov} and it is the only
 vacuum solution admitting a simply transitive $G_4$ as its maximal
group of motions.  This group of motions has two subgroups $G_3$ of
Bianchi Types $I$ and $VII_{h^2=\frac{1}{3}}$ acting in time-like
hyper-surfaces.
 \\ \\
$\blacktriangleright$ \textbf{Elementary solution of Painlev\'{e} transcendent}

As it is well known,although for generic values of the parameters
$(\alpha,\beta,\gamma,\delta)$ the Painlev\'{e} functions are
transcendental, there exist a lot of elementary solutions for
special values of these parameters \cite{Gromak},\cite{Newton}.
In the case at hand the following Lemma is applicable
\\ \\
\textbf{Lemma} \emph{The function $w$ satisfying $w(\xi)^2-2\,\xi\,w(\xi)+\xi=0$ is a solution of \eqref{Painleve 6} when the parameters  $(\alpha,\beta,\gamma,\delta)$ obey the relations $\alpha+\delta=\frac{1}{2},\, \beta=-\gamma$}.
\\ \\
\textbf{Proof} Direct computation. $\Box$
\\ \\

Using \eqref{par}, the conditions of the above Lemma are fulfilled
for $h=\pm\frac{2}{\sqrt{11}}$. Then from the first of\eqref{par} we
have
$(\alpha,\beta,\gamma,\delta)=(\frac{1}{8},\frac{11}{8},-\frac{11}{8},\frac{3}{8})$.
Choosing now the parametrization \be\label{ξ to τ}
w(\xi)=\frac{1}{4\,h^2}\,e^{4\,h\,\tau}, \quad
\xi=\cosh^2(2\,h\,\tau) \ee we can compute $y(\tau)$ from
\eqref{solution y} (with the minus sign) and $u_1(\tau),u_4(\tau)$
from \eqref{u1 and u4}, thereby arriving at the following line
element \be\label{Lucash sol}
d\,s^2&=&\kappa^2\,\Big(-e^{\frac{2\,\tau}{h}}\,\sinh^{-\frac{3}{8}}(4\,h\,\tau)\,(\bs{d}\,\tau)^2+
e^{-2\,h\,\tau}\,\sinh^{\frac{1}{2}}(4\,h\,\tau)\,(e^{4\,h\,\tau}+\sin(4\,\tau))\,(\bs{\sigma^1})^2
\non\\ && +
2\,e^{-2\,h\,\tau}\,\cos(4\,\tau)\,\bs{\sigma^1}\,\bs{\sigma^2}
+e^{-2\,h\,\tau}\,\sinh^{\frac{1}{2}}(4\,h\,\tau)\,(e^{4\,h\,\tau}-\sin(4\,\tau))\,(\bs{\sigma^2})^2
\non \\ &&
+e^{\frac{2\,\tau}{h}}\,\sinh^{-\frac{3}{8}}(4\,h\,\tau)\,(\bs{\sigma^3})^2
\Big)\ee  This geometry was first given by Lukash \cite{Lukash} and,
like \eqref{T3 VIIh}, admits only the three killing fields
\eqref{killing} and no homothetic vector field. Therefore, the
constant $\kappa$ is essential.

\section{Discussion}
In \cite{ChrTer JMP},\cite{ChrTer CQG} a systematic approach for
investigating the solution space of Bianchi Type Cosmologies  was
developed by the use of automorphisms and the theory of symmetries
of ordinary, coupled differential equations. The result was the
comprehensive recovery of all known closed form Type $III$ solutions,
as well as the presentation of the general solution in terms of the
sixth  Painlev\'{e} transcendent. In the present work we have
applied the method to the case of Bianchi Type $VII_h$ family of
vacuum geometries. Again, the general solution is implicitly given
in terms of the third \eqref{final h=0} Painlev\'{e} transcendent or
the  sixth Painlev\'{e} transcendent \eqref{final VII} for the Class
A ($h=0)$) and the Class B ($h\neq 0$) case respectively. Through
the investigation of either Particular or Elementary solutions of
the Painlev\'{e} transcendents we are able to concisely recover, in
a systematic fashion, all six known solutions \eqref{diagonal
h=0},\eqref{T3},\eqref{flat h},\eqref{G6 metric},\eqref{G4
h^2=1/3},\eqref{Lucash sol}. All these metrics have originally been
obtained in a time scale of 20 years or so,  by prior assumption of
symmetry and/or other physical requirements; e.g. Petrov' s solution
\cite{Petrov} was derived with the use of automorphisms  seeking
$G_4$ homogeneous metrics while Lukash' s solution \cite{Lukash} was
derived based on a physical interpretation of Type $VII_h$ cosmological
models, in terms of circularly polarized gravitational waves of
arbitrary wavelength in a space having constant negative curvature.
Their reacquisition single-handed, proves, we believe, the value of
our method. A very important result is, of course, the discovery of
the $\textbf{new}$ family of solutions \eqref{T3 VIIh} for the range
of the group parameter $h^2\leq\disfrac{1}{3}$. Besides of the obvious
value of a new family of solutions to the EFE's it also points to
the unexpected existence of a  sector with particular behavior for
this Bianchi Type. It is known that  Type $VI_h$ model has an
exceptional sector corresponding to the value $h^2=\disfrac{1}{9}$
but, for Type $VII_h$ such a behavior is first observed. The fact
may be taken as a further strengthening evidence of the widespread
belief that the two Types are very much similar. We hope that the
application of the method to Type $VI_h$ will bear analogous fruits.
As for Types $VIII, IX$, the recent discovery that some particular
configurations are described by the third Painlev\'{e} transcendent \cite{Conte}
strengthens our belief that their solution space will also be
attained by our method. We plan to return to these issues in the
immediate future. Finally, we deem it useful to end this discussion
by briefly describing the investigated solution space through the
following tables:
\newpage \thispagestyle{empty}
\begin{landscape}
\begin{center}
\begin{tabular}{ccc}
\multicolumn{3}{c}{{\Large\textbf{Bianchi Type $VII_0$ metrics}}} \\
\hline\hline \emph{Line Element} & \emph{Isometry Type} & \emph{Comments}\\
\hline & & \\
$d\,s^2=-(\bs{d}\,t)^2+(\bs{\sigma}^1)^2+(\bs{\sigma}^2)^2+
\tau^2\,(\bs{\sigma}^3)^2 $ & $G_{10}$ on $V_4$ & Flat Space\\ & & \\
\hline
\ph{$\disfrac{1}{\tau}$}& $G_4$ on $V_3$, $\tau>0$ & \\
$d\,s^2=-\tau\,(\bs{d}\,\tau)^2+\tau^2\,(\bs{\sigma}^1)^2+\tau^2\,(\bs{\sigma}^2)^2+
\disfrac{1}{\tau}\,(\bs{\sigma}^3)^2 $& $G_4$ on $T_3$, $\tau<0$ & LRS\\
& &\\
\hline
$d\,s^2=\kappa^2\,\left(e^{\tau^4}\,\tau\,(\bs{d}\,\tau)^2-\tau^2\,(\bs{\sigma}^1)^2+\tau^2\,(\bs{\sigma}^2)^2+
\disfrac{e^{\tau^4}}{\tau}\,(\bs{\sigma}^3)^2\right) $ & $G_3$ on $T_3$, $\tau>0$ & Non-homothetic\\
\hline & & \\
$d\,s^2=\kappa^2\,\left(-\disfrac{e^{u_1(\xi)}}{16\,\xi}\,(\bs{d}\,\xi)^2+\sqrt{\big|\xi\,w(\xi)\big|}\,(\bs{\sigma}^1)^2+
\sqrt{\Big|\disfrac{\xi}{w(\xi)}\Big|}\,(\bs{\sigma}^2)^2+
e^{u_1(\xi)}\,(\bs{\sigma}^3)^2 \right)$& $G_3$ on $V_3$ & General Solution \\
\hline
\end{tabular}
\end{center}
where the 1-forms $\bs{\sigma}^\alpha$ are given by \be
\bs{\sigma}^1=\sin x\,\bs{d}\,y+\cos x\,\bs{d}\,z,\,
\bs{\sigma}^2=\cos x \,\bs{d}\,y-\sin x\,\bs{d}\,z,\,
\bs{\sigma}^3=\frac{1}{2}\,\bs{d}\,x \ee and $u_1(\xi)$ is defined
by equation \eqref{u1'b}
\begin{eqnarray*}u_1'(\xi)&=\disfrac{\xi\,w'(\xi)^2}{4\,w(\xi)^2}
+\disfrac{1}{4}\,w(\xi)+
\disfrac{1}{4\,w(\xi)}-\frac{1}{4\,\xi}-\frac{1}{2}
\end{eqnarray*} with $w(\xi)$ standing for the third Painlev\'e
transcendent $w:=\mathbf{P_{III}}(-\frac{1}{2},\frac{1}{2},0,0)$,
defined by \eqref{eq w}

\newpage \thispagestyle{empty}
\begin{center}
\begin{tabular}{ccc}
\multicolumn{3}{c}{{\Large\textbf{Bianchi Type $VII_h$ metrics}}} \\
\hline\hline \emph{Line Element} & \emph{Isometry Type} & \emph{Comments}\\
\hline & & \\
$d\,s^2=-(\bs{d}\,\tau)^2+h^2\,\tau^2\,(\bs{\sigma}^1)^2+h^2\,\tau^2\,(\bs{\sigma}^2)^2+
4\,h^2\,\tau^2\,(\bs{\sigma}^3)^2  $ & $G_{10}$ on $V_4$ & Flat Space\\
\hline & & \\
$d\,s^2=\frac{1}{4}\,\exp\left(\frac{-2\,\lambda^2+2\,h^2\,(\lambda^2-1)}{h\,(\lambda^2-1)}\,\tau\right)
\left(-\bs{d}\,\tau^2+4\,(\bs{\sigma}^3)^2\right)+\frac{1}{2}\,e^{2\,h\,\tau}\,\left(1+\lambda\,\sin2\tau\right)\,(\bs{\sigma}^2)^2 $ & & \\
$-\frac{1}{2}\,e^{2\,h\,\tau}\,\left(-1+\lambda\,\sin2\tau\right)\,(\bs{\sigma}^1)^2+ e^{2\,h\,\tau}\,\lambda\,\cos2\tau\,(\bs{\sigma}^1)^2\,(\bs{\sigma}^2)^2$ & $G_6$ on $V_4$ & Homothetic\\
\hline && \\$d\,s^2=\kappa^2\,\sin(4\,h\,\tau) \Big(f(\tau)\,(\bs{d}\,\tau)^2 +\sin(h\,\ln f(\tau))\,(\bs{\sigma}^1)^2- \sin(h\,\ln f(\tau))\,(\bs{\sigma}^2)^2$
& & Non-Homothetic\\
$+ 2\,\cos(h\,\ln f(\tau))\,\bs{\sigma}^1\,\bs{\sigma}^2 +f(\tau)\, (\bs{\sigma}^3)^2\Big)$ & $G_3$ on $T_3$ & $|h|\leq \frac{1}{\sqrt{3}}$ \\
 \hline && \\
 $d\,s^2= \kappa^2\Big(\csc^2\frac{4\,\tau}{\sqrt{3}}(\bs{d}\,\tau)^2 -\sin\frac{4\,\tau}{\sqrt{3}}\,\sin\big(\sqrt{3}\,\ln\sin\frac{4\,\tau}{\sqrt{3}}\big)\,(\bs{\sigma}^1)^2$ & & Non-Homothetic\\ $+\,\sin\frac{4\,\tau}{\sqrt{3}}\,\sin\big(\sqrt{3}\,\ln\sin\frac{4\,\tau}{\sqrt{3}}\big)\,(\bs{\sigma}^2)^2$ & & Maximal $G_4$\\
$+2\,\sin\frac{4\,\tau}{\sqrt{3}}\,\cos\big(\sqrt{3}\,\ln\sin\frac{4\,\tau}{\sqrt{3}}\big)\,\bs{\sigma}^1\,\bs{\sigma}^2 +\csc^2\frac{4\,\tau}{\sqrt{3}}(\bs{\sigma}^3)^2\Big)$ & $G_4$ on $T_3$ & $h^2=\frac{1}{3}$\\ \hline && \\ $d\,s^2=\kappa^2\,\Big(-e^{\frac{2\,\tau}{h}}\,\sinh^{-\frac{3}{8}}(4\,h\,\tau)\,(\bs{d}\,\tau)^2+ e^{-2\,h\,\tau}\,\sinh^{\frac{1}{2}}(4\,h\,\tau)\,(e^{4\,h\,\tau}+\sin(4\,\tau))\,(\bs{\sigma^1})^2$ &&\\
$+ 2\,e^{-2\,h\,\tau}\,\cos(4\,\tau)\,\bs{\sigma^1}\,\bs{\sigma^2} +e^{-2\,h\,\tau}\,\sinh^{\frac{1}{2}}(4\,h\,\tau)\,(e^{4\,h\,\tau}-\sin(4\,\tau))\,(\bs{\sigma^2})^2$ && Non-Homothetic\\ $+e^{\frac{2\,\tau}{h}}\,\sinh^{-\frac{3}{8}}(4\,h\,\tau)\,(\bs{\sigma^3})^2 \Big)$ & $G_3$ on $V_3$ & $h^2=\frac{4}{11}$\\ \hline && \\
$d\,s^2=\kappa^2\,\Bigg(-\frac{e^{u_1(\xi)}}{16\,h^2\,\xi\,(\xi-1)}\,(\bs{d}\,\xi)^2+ \sqrt{\xi\,(\xi-1)}\left(\sqrt{y'(\xi)}- \sin(2\,u_4(\xi))\,\sqrt{y'(\xi)-\frac{1}{h^2}}\right)\,(\bs{\sigma}^1)^2 $ && \\ $+2\,\cos(2\,u_4(\xi))\,\sqrt{\xi\,(\xi-1)\,(y'(\xi)-\frac{1}{h^2})}\,\bs{\sigma}^1\,\bs{\sigma}^2$ && \\$+ \sqrt{\xi\,(\xi-1)}\left(\sqrt{y'(\xi)}+ \sin(2\,u_4(\xi))\,\sqrt{y'(\xi)-\frac{1}{h^2}}\right)\,(\bs{\sigma}^2)^2+ e^{u_1(\xi)}\,(\bs{\sigma}^3)^2\Bigg)$ & $G_3$ on $V_3$ & General Solution\\ \hline
\end{tabular}
\end{center}
\end{landscape} where the 1-forms $\bs{\sigma}^\alpha$ are given by \be
\bs{\sigma}^1=e^{h\,x}\left(\sin x\,\bs{d}\,y+\cos x\,\bs{d}\,z\right),\,
\bs{\sigma}^2=e^{h\,x}\left(\cos x \,\bs{d}\,y-\sin x\,\bs{d}\,z\right),\,
\bs{\sigma}^3=\frac{1}{2}\,\bs{d}\,x \ee  the function $f(\tau)$ stands for
\be f(\tau) =  \sin^{-\frac{1}{h^2}}(4\,h\,\tau)\, \tan^{-\frac{\sqrt{1-3\,h^2}}{h^2}}(2\,h\,\tau)\ee and the functions $u_1(\xi),\,u_4(\xi)$ are defined by \eqref{u1 and u4}
\bsub\be u_1'(\xi) & = & \frac{\left( -1 + h^2 \right) \,\left( -1 + 2\,\xi  \right)  + 2\,h^2\,y(\xi )}
  {2\,h^2\,\left( -1 + \xi  \right) \,\xi } \ee
\be u_4'(\xi) & = & \frac{1 - 2\,\xi  + 2\,h^2\,y(\xi )}
  {4\,h\,\left( -1 + \xi  \right) \,\xi \,\left( -1 + h^2\,y'(\xi ) \right) }\ee \esub
with $y(\xi)$ defined by \eqref{solution y}.

%%%%%%%%%%%%%%%%%%%%%%%%%%%%%%%%%%%%%%%%%%%%%%%%%%%%%%%%%%%%%%%%%%%%%%%%%%%%%%%%%%%%%%%%

\end{document}